\documentclass[twocolumn, preprintnumbers,amsmath,amssymb,balancelastpage]{revtex4}
\pdfoutput=1

\usepackage{putexPRL}
\usepackage{graphicx}
\usepackage{dsfont}
\usepackage{amsmath}
\usepackage{array}
\usepackage{epstopdf}
\usepackage{enumerate}
\usepackage{youngtab}
\usepackage{tensor}
\usepackage{braket}
\usepackage[normalem]{ulem}
\usepackage{slashed}
\usepackage[aligntableaux=center]{ytableau}
\usepackage[utf8]{inputenc}
\usepackage[
      colorlinks=true,
      linkcolor=blue,
      urlcolor=blue,
      filecolor=black,
      citecolor=red,
      ]{hyperref}
\usepackage{braket}
\usepackage{mathtools}
\usepackage{rotating}
\usepackage{tabularx}
\newcolumntype{Y}{>{\centering\arraybackslash}X}
\newcolumntype{C}[1]{>{\centering\arraybackslash}p{#1}}
\usepackage{todonotes}
\usepackage{xcolor}
\definecolor{LightCyan}{rgb}{0.7,1,1}
\definecolor{Gray}{gray}{0.9}
\usepackage{xspace}

\newcommand {\be} {\begin {equation}}
\newcommand {\ee} {\end {equation}}

\newcommand {\bes} {\begin {equation*}}
\newcommand {\ees} {\end {equation*}}

\newcommand{\es}[2] {\begin{equation} \label{#1} \begin{split} #2 \end{split} \end{equation}}

\newcommand{\cO}{{\mathcal O}}

\newcommand{\beq}{\begin{equation}}
\newcommand{\eeq}{\end{equation}}

\def\ie{\begin{equation}\begin{aligned}}
\def\fe{\end{aligned}\end{equation}}

\def\<{\langle}
\def\>{\rangle}

\def\beg{\begin{equation}\begin{gathered}}
\def\eeg{\end{gathered}\end{equation}}

\def\bea{\begin{equation}\begin{aligned}}
\def\eea{\end{aligned}\end{equation}}

\begin{document}

\title{Upper critical dimension of the 3-state Potts model}

\author{Shai M.~Chester}         
\affiliation{
Jefferson Physical Laboratory, Harvard University, Cambridge, MA 02138, USA\\
Center of Mathematical Sciences and Applications, Harvard University, Cambridge, MA 02138, USA
		}
\author{Ning Su}
\affiliation{Department of Physics, University of Pisa, I-56127 Pisa, Italy}

\begin{abstract}
We consider the 3-state Potts model in $d\geq2$ dimensions. For $d$ less than the upper critical dimension $d_\text{crit}$, the model has a critical and a tricritical fixed point. In $d=2$, these fixed points are described by minimal models, and so are exactly solvable. For $d>2$, however, strong coupling makes them difficult to study and there is no consensus on the value of $d_\text{crit}$. We use the numerical conformal bootstrap to compute critical exponents of both the critical and tricritical fixed points for general $d$. In $d=2$ our results match the expected values, and as we increase $d$ we find that the critical exponents of each fixed point get closer until they merge near $d_\text{crit}\lesssim 2.5$. 
\end{abstract}

\maketitle
\nopagebreak

\section{Introduction}

Lattice models in $d$ dimensions \footnote{In condensed matter language, this would be $(d-1)+1$ dimensions.} are useful descriptions of many physical systems such as magnets and superfluids. For a range of $d$, these models can be tuned to undergo an interacting second order phase transition, which is described by a unitary conformal field theory (CFT). For instance, the Ising model has a phase transition described by a CFT in $2\leq d\leq4$, but in $d\geq4$ the CFT becomes a free theory. The largest value of $d$ such that a theory flows to an interacting unitary CFT is called the upper critical dimension. Surprisingly, the upper critical dimension is still unknown for the simplest lattice model after the Ising model: the 3-state Potts model. We will address this question using the modern conformal bootstrap.

The $q$-state Potts model is defined on a square lattice of random spins $s_i$ by the partition function \cite{potts_1952}
\es{Z}{
Z=\sum_{\{s_i\}}e^{-H[\{s_i\}]}\,,\qquad H[\{s_i\}]=  \beta\sum_{\langle ij\rangle}\delta_{s_i,s_j}\,,
}
where $\beta$ is the inverse temperature. At large temperature this model has a disordered phase with one ground state with $S_q$ symmetry, while at small temperature $S_q$ is broken and one spin value is preferred. We can tune $\beta=\beta_\text{crit}$ to get a phase transition called the critical Potts model. If we allow some lattice sites to be vacant \footnote{The model with vacancies is called the dilute Potts model.}, we can tune the chemical potential of these vacancies along with $\beta$ to get another transition called the tricritical Potts model. In $d=2$, both the tricritical and critical phase transitions are second order for $q\leq q_\text{crit}=4$ \cite{Baxter_1973} \footnote{The $q=4$ theory is described by unitary conformal field theory given by a free scalar compactified on $S_1/\mathbb{Z}_2$ with radius $R=1/\sqrt{2}$ \cite{Dijkgraaf:1987vp}.}, and so are described by a conformal field theory. In fact, as $q\to q_\text{crit}$ the critical exponents of each theory merge and then go off into the complex plane \cite{PhysRevLett.43.737,Nienhuis_1980,Gorbenko:2018dtm,Gorbenko:2018ncu}, which is an example of the merger and annihilation scenario of CFTs \cite{PhysRevD.80.125005}. 

The $q=3$ model is of special experimental and theoretical interest for general $d$. The critical model has applications in $d=2$ to ${}^4$He atoms on graphite \cite{PhysRevLett.44.152,PhysRevLett.38.501,ALEXANDER1975353}. In $d=2$, the critical and tricritical CFTs are described by minimal models, and so are exactly solvable \cite{BELAVIN1984333,PhysRevLett.52.1575}. In $d\sim6$ one can write down weakly coupled Landau-Ginzburg Lagrangians for these theories using a complex field \cite{Zia:1975ha,Amit_1976}, but this description is very strongly coupled in the $2\leq d<3$ regime of interest  \footnote{Another perturbative approach was proposed using an expansion around $d=4-\epsilon$ and $q=2+\epsilon$ \cite{PhysRevB.23.362}, but it was shown in \cite{Newman:1984hy} that the theory is not perturbative in this regime.}, where we require non-perturbative methods. In $d=3$, lattice Monte Carlo simulations suggest that the $q=3$ critical model is first order, so there is no CFT description, and in fact $q_\text{crit}\sim2.45$ \cite{PhysRevB.43.1268} \footnote{Fractional values of $q$ can be made sense of using the random-cluster definition of the Potts model \cite{FORTUIN1972536}.}. For $2<d<3$, the upper critical dimension $d_\text{crit}$ for $q=3$ has not been determined, and it is not even known if the critical and tricritical models disappear via the merger and annihilation scenario \footnote{In $d>2$, it is a priori possible that the tricritical model could annihilate with the gaussian free theory \cite{PhysRevB.23.6055}.}. Lattice simulations in fractional dimensions give predictions for $(d_\text{crit},q_\text{crit})$ that are listed in Table \ref{tab:table1}, but there is no consensus for $d_\text{crit}$ when $q=3$. It is also not clear if the fractional $d$ lattices used in these studies are analytically connected to the integer $d$ fixed points, as indeed some of the different fractional $d$ studies in Table \ref{tab:table1} give conflicting answers in $d=3$.

\begin{table}
\begin{ruledtabular}
\begin{tabular}{c|cccccccc}
$(d_\text{crit},q_\text{crit}):$& (2,4)& $(2.32,2.85)$ & $(2.5,2.68)$  & $(3,2.45)$
 \\
&&  & $(3,2.21)$  & 
 \\
Ref.& \cite{Baxter_1973} & \cite{PhysRevB.23.6055} & \cite{PhysRevA.44.8000} & \cite{PhysRevB.43.1268}  \\
\end{tabular}
\end{ruledtabular}
\caption{\label{tab:table1} Previous estimates for the upper critical dimension $d_\text{crit}$ for various $q$. For $d=2$ the result is exact, while the $d>2$ cases come from lattice studies.
 }
\end{table}

In this paper we will use the numerical conformal bootstrap \cite{Rattazzi:2008pe} \footnote{See \cite{Rychkov:2016iqz,Simmons-Duffin:2016gjk,Poland:2018epd,Chester:2019wfx,Poland:2022qrs} for reviews.} to estimate $d_\text{crit}$ for $q=3$. This method bounds the allowed values of critical exponents for any CFT with a given symmetry, by applying a finite truncation of the infinite constraints of crossing equations for a given set of four-point functions. Special features in the space of the allowed region often correspond to physical theories. For instance, a kink in the space of allowed critical exponents of $O(N)$ invariant theories was found to match Monte Carlo predictions for the critical $O(N)$ models \cite{ElShowk:2012ht,Kos:2013tga}, and this correspondence was later confirmed by more sophisticated bootstrap studies that restricted the allowed region to a tiny island around the kink \cite{Kos:2014bka,Kos:2016ysd,Chester:2019ifh,PhysRevD.104.105013}.

To bootstrap the 3-state critical Potts model, we consider correlations functions of the two $S_3$ charged operators and the one singlet operator that were relevant in the known $d=2$ theory \footnote{See \cite{Rong:2017cow} for an earlier study numerical bootstrap study of 2d CFTs with $S_3$ symmetry.}. We find that the allowed region in the space of the critical exponents of these three operators is roughly approximated by a cone for each $d$. The numerics are extremely intensive, as they require scanning over not only the critical exponents, but also their various three-point structures (i.e. OPE coefficients). To efficiently find the tip of the cone, we use the recently developed Navigator method \cite{Reehorst:2021ykw} to minimize the lowest lying charged relevant critical exponent, which is much faster than exploring the full allowed region.  In $d=2$, the cone tip precisely matches the known minimal model values, which motivates our conjecture that the tip in $d>2$ continues to describe the physical theory. The bootstrap setup for the tricritical theory is similar, except there is now an extra relevant singlet operator, so we scan over a four-dimensional space and again find a four-dimensional generalization of the tip in general $d$ that matches the known minimal model in $d=2$. As we increase $d$ above $2$, we find that the tips identified with the critical and tricritical theories get closer to each other until around $d\sim 2.5$, where the theories merge, and in particular the extra singlet operator in the tricritical theory becomes marginal. Our results are summarized by Figure \ref{fig:spectrum}, and we will discuss two possibilities for $d_\text{crit}\lesssim2.5$ based on this data.

The rest of this paper is organized as follows.  In Section~\ref{sec:review}, we review properties of the conformal field theories that describe the critical and tricritical fixed points in general $d$, starting with the exactly known $d=2$ minimal model descriptions. In Section~\ref{sec:bootstrap}, we describe our bootstrap setup, how to numerically implement it using the Navigator, and the resulting estimates for the critical exponents. We end with a discussion of our results in Section~\ref{sec:discussion}.

\section{Critical and tricritical CFTs}\label{sec:review}

We begin by reviewing the CFTs that describe the critical and tricritical fixed points of the 3-state Potts model. Operators in these theories transform in representations of the global conformal symmetry group $SO(d,2)$, as labeled by scaling dimension $\Delta$ and spin $\ell$, as well as representations $\bf r$ of the flavor group $S_3$. The $S_3$ symmetry group has three irreducible representations: the dimension two fundamental $\bf 1$ (i.e. the charged), the dimension 1 singlet $\bf0^+$, and the dimension 1 sign representation $\bf 0^-$. By definition, the critical theory has one relevant spin zero singlet $\epsilon$, while the tricritical has an extra relevant spin zero singlet $\epsilon'$. As unitary local CFTs, they must also contain a conserved stress tensor operator with $\Delta=d$ and spin 2. A priori, this is all we know about these CFTs in general $d\geq2$.

For $d=2$, the global conformal symmetry group $SO(d,2)$ is enhanced to a pair of infinite dimensional Virasoro groups $\mathcal{V}_\text{l}\otimes \mathcal{V}_\text{r}$. This enhanced symmetry group was used in \cite{BELAVIN1984333, PhysRevLett.52.1575} to solve a set of theories known as Virasoro minimal models $\mathcal{M}_{p,q}$ labeled by coprime integers $q>p>2$ and central charge
\es{c}{
c_{p,q}=1-6\frac{(p-q)^2}{pq}\,.
}
The partition functions of these CFTs must also be modular invariant, and there are different ways of imposing this constraint for a given $(p,q)$, which yield a different spectrum of operators and thus a different theory. When $q=p+1$ and $p<5$, there is a unique invariant that requires the weights $h_{l,r}$ under $\mathcal{V}_\text{l,r}$ to be identical. These theories are the free theory for $p=2$, and the $p$-critical Ising model for $p>2$, e.g. $p=2$ is the critical Ising, $p=3$ is the tricritical Ising, etc. These minimal models have a $\mathbb{Z}_2$ invariant Landau-Ginzburg Lagrangian in terms of a real scalar field $\phi$ with potential $\phi^{2(p-1)}$. The upper critical dimension of these theories can thus be trivially determined by demanding that $\phi^{2(p-1)}$ is marginal, i.e. that the theory merge and annihilate with the free theory, which fixes \footnote{These upper critical dimensions have also been verified with the numerical bootstrap in \cite{Gowdigere:2018lxz}.}
\es{dcriteasy}{
d_\text{crit}^{p}=2\frac{p-1}{p-2}\,.
}
For instance, for the critical Ising model with $p=3$, we recover the expected $d_\text{crit}^3=4$.

For $p\geq5$, there is now a non-diagonal modular invariant where $h_{l,r}$ can be different. For $p=5$ this yields the critical Potts model, with Virasoro primaries \footnote{Note the spin three generator $W$, which enhances the Virasoro symmetry to $\mathcal{W}_3$, such that the critical 3-state Potts model is the lowest $c$ member of an infinite class of $\mathcal{W}_3$ minimal models \cite{Fateev:1987vh}.}
\es{critMin}{
\text{I} &=[0,0,{\bf0}^+],\;\;\;\;\;\; \sigma=[{2}/{15},0,{\bf 1}],\;\; \sigma'=[{4}/{3},0,{\bf 1}]\,,\\ 
\varepsilon&=[{4}/{5},0,{\bf 0}^+],\; \;\varepsilon'=[{14}/{5},0,{\bf 0}^+],\;\;\varepsilon_2=[6,0,{\bf 0}^+]\,,\\ 
\gamma&=[9/{5},1,{\bf 0^-}],\quad W=[3,3,{\bf 0^-}]\\ 
}
labelled as $[\Delta,\ell, {\bf r}]$, and these Virasoro primaries generate an infinite set of quasiprimaries under the global conformal group $SO(2,2)$, such as the stress tensor. Note that there is only one relevant singlet scalar, $\varepsilon$, as expected for a critical fixed point, there are two relevant charged scalars $\sigma$ and $\sigma'$, and no relevant scalars in the $\bf0^-$. The OPE coefficients have also been computed and are given in \cite{1995cond.mat..7033M}, but we will not make use of them. For $p=6$, the non-diagonal minimal model is the tricritical Potts model, with Virasoro primaries \footnote{Note the spin five generator $W$, which enhances the Virasoro symmetry to $\mathcal{W}(2,5)$. This is the only known CFT with $\mathcal{W}(2,5)$ symmetry, since this symmetry is exceptional and can only exist for finitely many values of $c$, unlike $\mathcal{W}_3$ which can exist for infinitely many values \cite{Bouwknegt:1992wg}. On the other hand, the tricritical theory is the lowest $c$ member of a family of unitary CFTs with $S_3$ symmetry built from parafermions \cite{Fateev:1985ig}.}
\es{tricritMin}{
\hspace{-.1in}\text{I} &=[0,0,{\bf0}^+], \;\;\sigma=[{2}/{21},0,{\bf 1}],\;\;\sigma'=[{20}/{21},0,{\bf 1}], \\ 
\hspace{-.1in}\sigma_2&=[{8}/{3},0,{\bf 1}],\;\; \varepsilon=[{2}/{7},0,{\bf 0}],\;\;\varepsilon'=[{10}/{7},0,{\bf 0}],\\ 
\hspace{-.1in} \varepsilon_2&=[{24}/{7},0,{\bf 0}^+], \varepsilon_3=[{44}/{7},0,{\bf 0}^+], \varepsilon_4=[10,0,{\bf 0}^+],\\ 
\hspace{-.1in}\gamma&=[17/{7},1,{\bf 0^-}],\;\gamma'=[23/{7},3,{\bf 0^-}],\; W=[5,5,{\bf 0^-}].\\ 
}
There are now two relevant singlets, $\varepsilon$ and $\varepsilon'$, as we expect for a tricritical fixed point, as well as two relevant charged scalars $\sigma$ and $\sigma'$, and still no relevant scalars in the $\bf0^-$. Some OPE coefficients have also been computed in \cite{Fateev:1985ig}.

\section{Numerical conformal bootstrap}\label{sec:bootstrap}

We will now describe our numerical bootstrap study of the critical and tricritical CFTs in $d\geq2$. In this section we will only consider the global conformal group $SO(d,2)$ even for $d=2$, so that our setup applies the same way in any $d\geq2$. 

\subsection{Crossing equations}\label{cross}

Recall that the $S_3$ flavor group has three irreducible representations: $\bf1$, $\bf0^+$, and $\bf0^-$. The tensor products of these representations are
\es{prods}{
\bf 1\otimes \bf 1&={\bf 1}_s\oplus {\bf0}^+_s\oplus {\bf0}^-_a\,,\qquad\bf 1\otimes \bf0^\pm = \bf 1\,,\\ 
\bf 0^\pm\otimes \bf 0^\mp&=\bf 0^-\,,\qquad \qquad\quad\;\; \, \bf 0^\pm\otimes \bf0^\pm = \bf0^+\,,\\ 
}
where $\bf s/a$ denotes the symmetric/antisymmetric product. Four-point functions of scalar operators $\varphi_{\bf r}(x)$ in irrep $\bf r$ can be expanded in the $s$-channel in terms of $SO(d,2)$ conformal blocks $g^{\Delta^-_{12},\Delta^-_{34}}_{\Delta,\ell}(u,v)$ \cite{Dolan:2003hv} as
\es{4point}{
&\left\langle  \varphi_{{\bf r}_1}(x_1)  \varphi_{\bf r_2}(x_2)   \varphi_{\bf r_3}(x_3)   \varphi_{\bf r_4}(x_4)  \right\rangle=  \frac{x^{\Delta^-_{12}}_{24} x^{\Delta^-_{34}}_{14}}{x^{\Delta^-_{12}}_{14} x_{13}^{\Delta^-_{34}} }      \\
&\times\frac{1} {x_{12}^{\Delta_{12}^+}x_{34}^{\Delta_{34}^+}} \sum_{\cO}\lambda^\cO_{\varphi_1\varphi_2}\lambda^\cO_{\varphi_3\varphi_4}T^{\bf r}_{\bf r_1\bf r_2\bf r_3\bf r_4}g^{\Delta^-_{12},\Delta^-_{34}}_{\Delta,\ell}(u,v),
}
 where $\Delta^\pm_{ij}\equiv\Delta_i\pm\Delta_j$, the conformal cross ratios $u,v$ are 
 \es{uv}{
   u \equiv \frac{{x_{12}^2x_{34}^2}}{{x_{13}^2x_{24}^2}},\qquad v \equiv \frac{{x_{14}^2x_{23}^2}}{{x_{13}^2x_{24}^2}}\,,
 }
and the operators $\cO$ that appear in both OPEs $\varphi_1\times\varphi_2$ and $\varphi_3\times\varphi_4$ have scaling dimension $\Delta$, spin $\ell$, and transform in an irrep $\bf r$ that appears in both the tensor products ${\bf r}_1\otimes{\bf r}_2$ and ${\bf r}_3\otimes{\bf r}_4$. If $\varphi_1=\varphi_2$ (or $\varphi_3=\varphi_4$), then Bose symmetry requires that $\cO$ have only even/odd $\ell$ for ${\bf r}$ in the symmetric/antisymmetric product of ${\bf r}_1\otimes{\bf r}_2$ (or ${\bf r}_3\otimes{\bf r}_4$). Equating this $s$-channel expansion with the $t$-channel expansion, given by swapping $\varphi_{{\bf r}_1}(x_1)$ and $\varphi_{{\bf r}_3}(x_3)$, gives the crossing equations
\es{crossing}{
0=& \sum_{\cO}\lambda^\cO_{\varphi_1\varphi_2}\lambda^\cO_{\varphi_3\varphi_4}T^{\bf r}_{\bf r_1\bf r_2\bf r_3\bf r_4}v^{\frac{\Delta_{23}^+}{2}}g^{\Delta^-_{12},\Delta^-_{34}}_{\Delta,\ell}(u,v)\\
&-\sum_{\cO}\lambda^\cO_{\varphi_3\varphi_2}\lambda^\cO_{\varphi_1\varphi_4}T^{\bf r}_{\bf r_3\bf r_2\bf r_1\bf r_4}u^{\frac{\Delta_{12}^+}{2}}g^{\Delta^-_{32},\Delta^-_{14}}_{\Delta,\ell}(v,u)\,,
}
which can be further decomposed into a finite set of equations as a function of $u,v$ using the explicit form of the tensor structure $T^{\bf r}_{\bf r_1\bf r_2\bf r_3\bf r_4}$. In table \ref{configs} we list the 4-point functions of $\sigma$, $\sigma'$, and $\epsilon$ that are allowed by $S_3$ symmetry and whose $s$ and $t$-channel configurations lead to independent crossing equations, along with the irreps and spins of the operators that appear in the OPE, and the number of crossing equations that they yield. The explicit crossing equations can be automatically generated using \texttt{autoboot} \cite{Go:2019lke}, whose conventions we use for the tensor structures and the conformal blocks. The explicit crossing equations can be found in the attached \texttt{Mathematica} file. 

\begin{table}
\begin{ruledtabular}
\begin{tabular}{c|c|c|c}
 Correlator& $s$-channel & $t$-channel&Eqs\\
 \hline 
$\langle\sigma\sigma\sigma\sigma\rangle$&  ${\bf0^+_s}$, ${\bf0^-_a}$, ${\bf1_s}$&same&  3   \\
 \hline
 $\langle\sigma'\sigma'\sigma'\sigma'\rangle$&  ${\bf0^+_s}$, ${\bf0^-_a}$, ${\bf1_s}$&same&  3   \\
 \hline
 $\langle\sigma\sigma'\sigma\sigma'\rangle$&  ${\bf0^+}$, ${\bf0^-}$, ${\bf1}$&same&  3   \\
 \hline
 $\langle\sigma\sigma\sigma'\sigma'\rangle$&  ${\bf0^+_s}$, ${\bf0^-_a}$, ${\bf1_s}$&${\bf0^+}$, ${\bf0^-}$, ${\bf1}$&  6   \\
 \hline
 $\langle\sigma\sigma'\sigma'\sigma'\rangle$&  ${\bf0^+_s}$, ${\bf0^-_a}$, ${\bf1_s}$& same &  3   \\
 \hline
 $\langle\sigma'\sigma\sigma\sigma\rangle$&  ${\bf0^+_s}$, ${\bf0^-_a}$, ${\bf1_s}$& same &  3   \\
 \hline
 $\langle\varepsilon\varepsilon\varepsilon\varepsilon\rangle$&  ${\bf0^+_s}$ & same &  1   \\
 \hline
 $\langle\varepsilon\sigma\varepsilon\sigma\rangle$&  ${\bf1}$ & same &  1   \\
 \hline
  $\langle\varepsilon\sigma'\varepsilon\sigma'\rangle$&  ${\bf1}$ & same &  1   \\
 \hline
   $\langle\varepsilon\sigma\varepsilon\sigma'\rangle$&  ${\bf1}$ & same &  1   \\
 \hline
 $\langle\varepsilon\varepsilon\sigma\sigma\rangle$&  ${\bf0^+_s}$ &  ${\bf1}$ &  2   \\
 \hline
$\langle\varepsilon\varepsilon\sigma'\sigma'\rangle$&  ${\bf0^+_s}$ &  ${\bf1}$ &  2   \\
 \hline
$\langle\varepsilon\varepsilon\sigma\sigma'\rangle$&  ${\bf0^+_s}$ &  ${\bf1}$ &  2   \\
 \hline
$\langle\varepsilon\sigma\sigma\sigma\rangle$&   ${\bf1_s}$  &  same &  1   \\
\hline
$\langle\varepsilon\sigma'\sigma'\sigma'\rangle$&   ${\bf1_s}$  &  same &  1   \\
\hline
$\langle\varepsilon\sigma\sigma'\sigma'\rangle$&  ${\bf1_s}$  &  ${\bf1}$ &  2   \\
\hline
$\langle\varepsilon\sigma'\sigma\sigma\rangle$&  ${\bf1_s}$ &  ${\bf1}$ &  2   \\
\hline
$\langle\sigma\varepsilon\sigma\sigma'\rangle$&  ${\bf1}$ &  same&  1   \\
\hline
$\langle\sigma'\varepsilon\sigma'\sigma\rangle$&  ${\bf1}$ &  same&  1   \\
\end{tabular}
\end{ruledtabular}
\caption{Four-point function configurations that give independent crossing equations under equating their $s$- and $t$-channels, where $\bf s/a$ denote that only even/odd spins appear.}
\label{configs}
\end{table}

\begin{figure}
	\centering
	\includegraphics[width=\columnwidth]{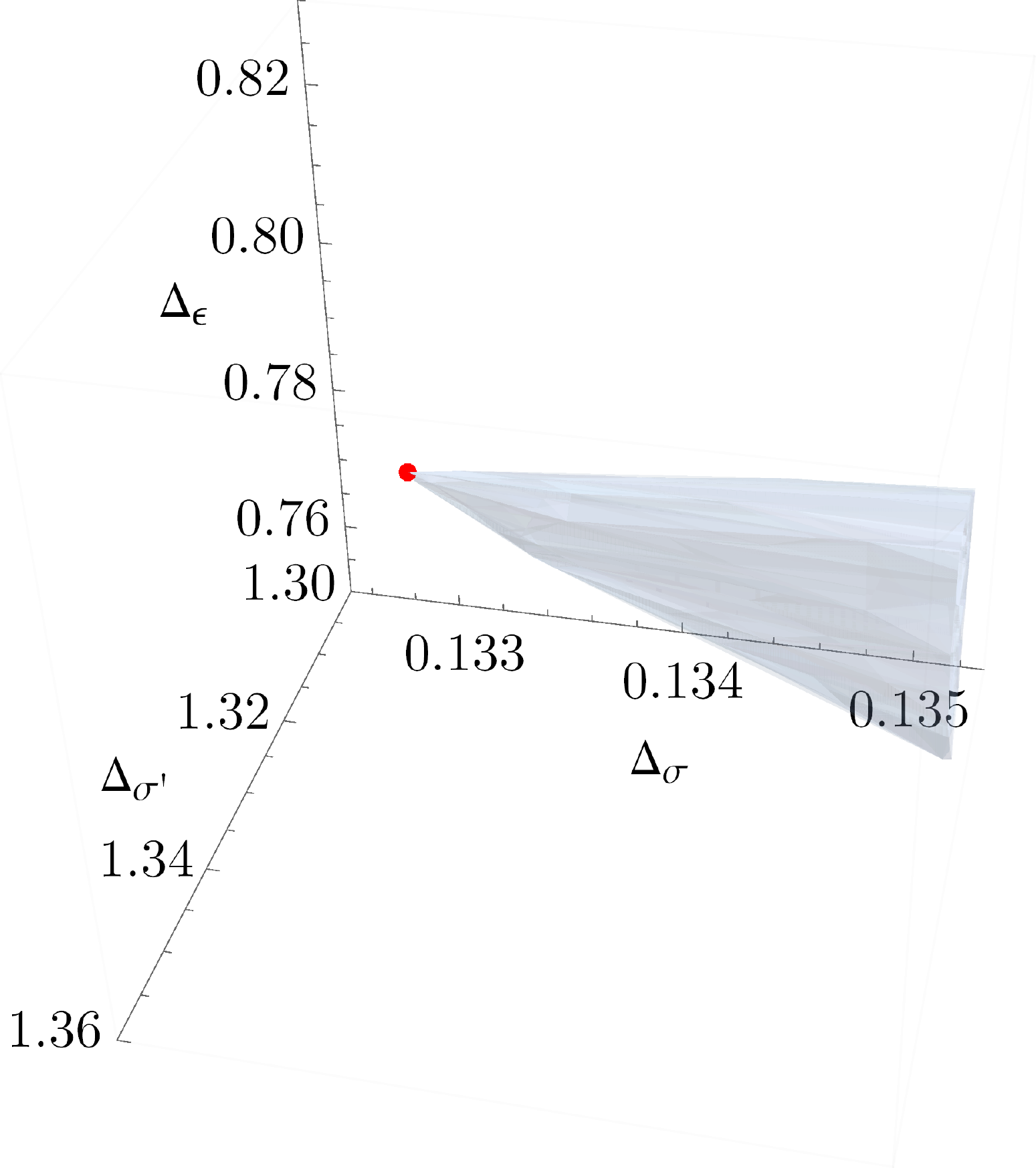}
		\caption{The cone-shaped allowed region in the space of $(\Delta_\sigma,\Delta_{\sigma'},\Delta_\epsilon)$ in $d=2$, obtained by bootstrapping correlators of $\sigma,\sigma'\epsilon$ and assuming they are the only relevant scalars in their sector. The red dot corresponds to the exact solution of the critical Potts model, which matches the tip of the cone.}
	\label{fig:2d}
\end{figure}

\subsection{Numerical implementation}
\label{nav}

\begin{figure*}
	\centering
	\includegraphics[width=\columnwidth]{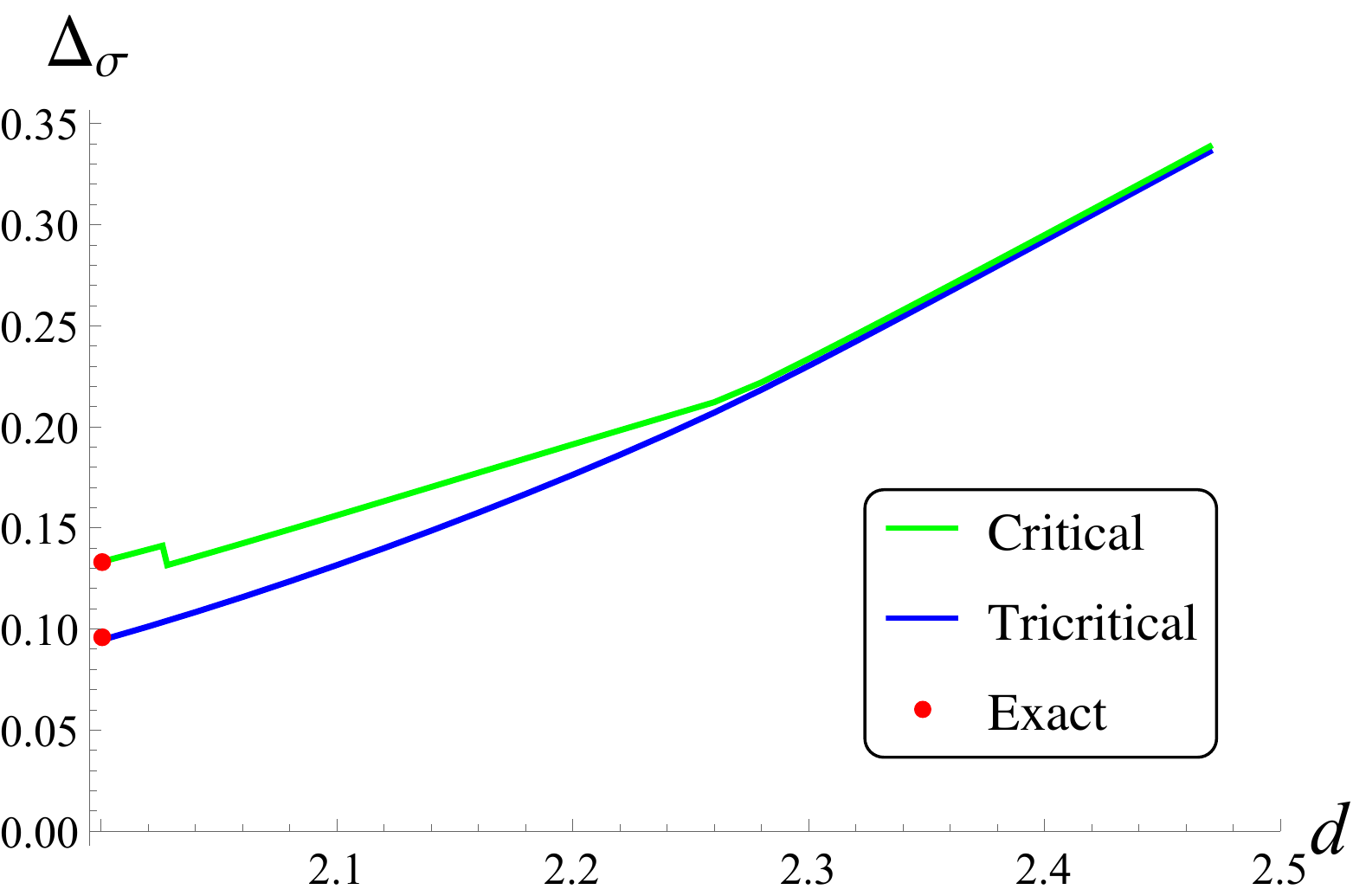}
	\includegraphics[width=\columnwidth]{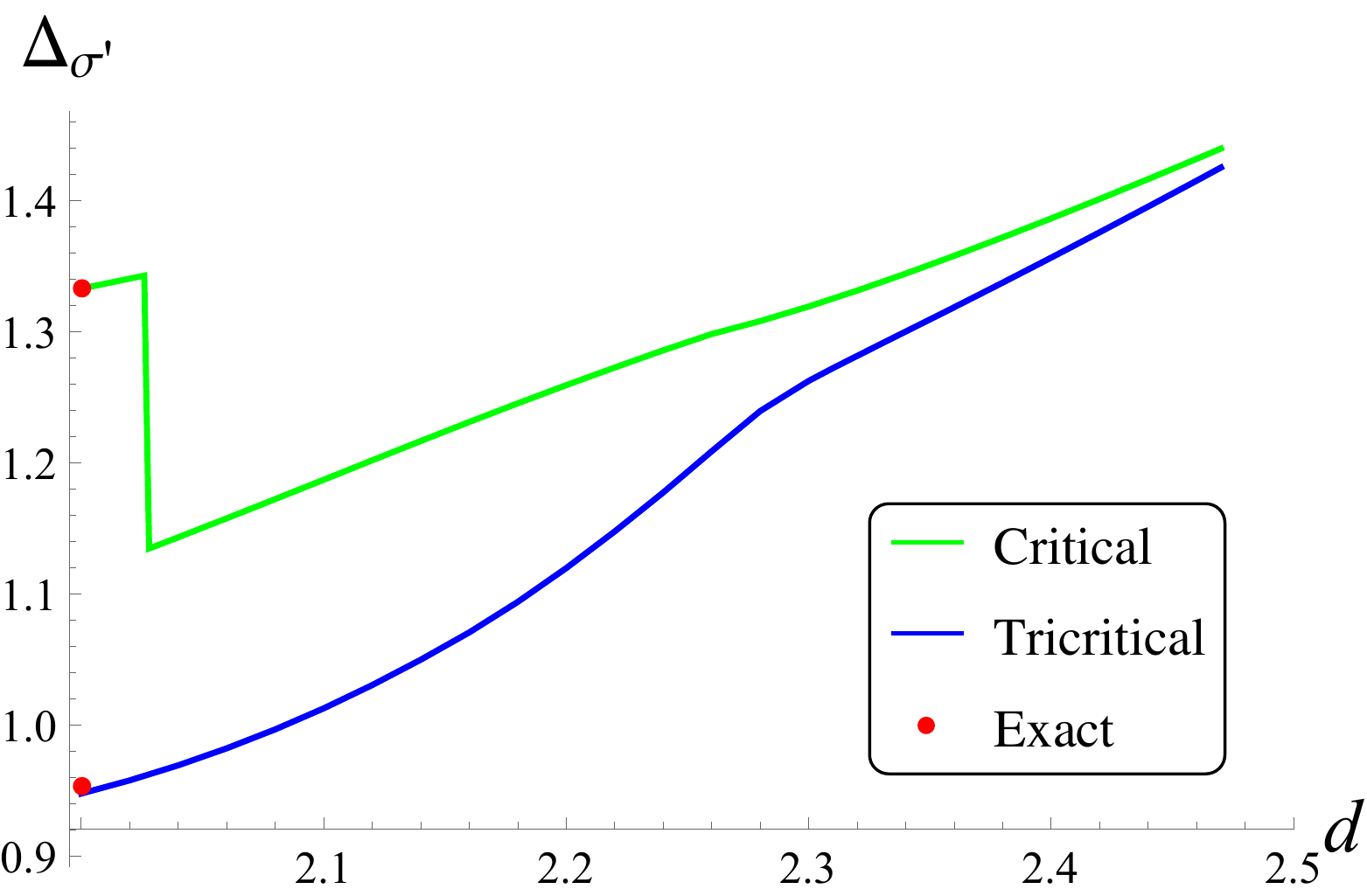}
	\includegraphics[width=\columnwidth]{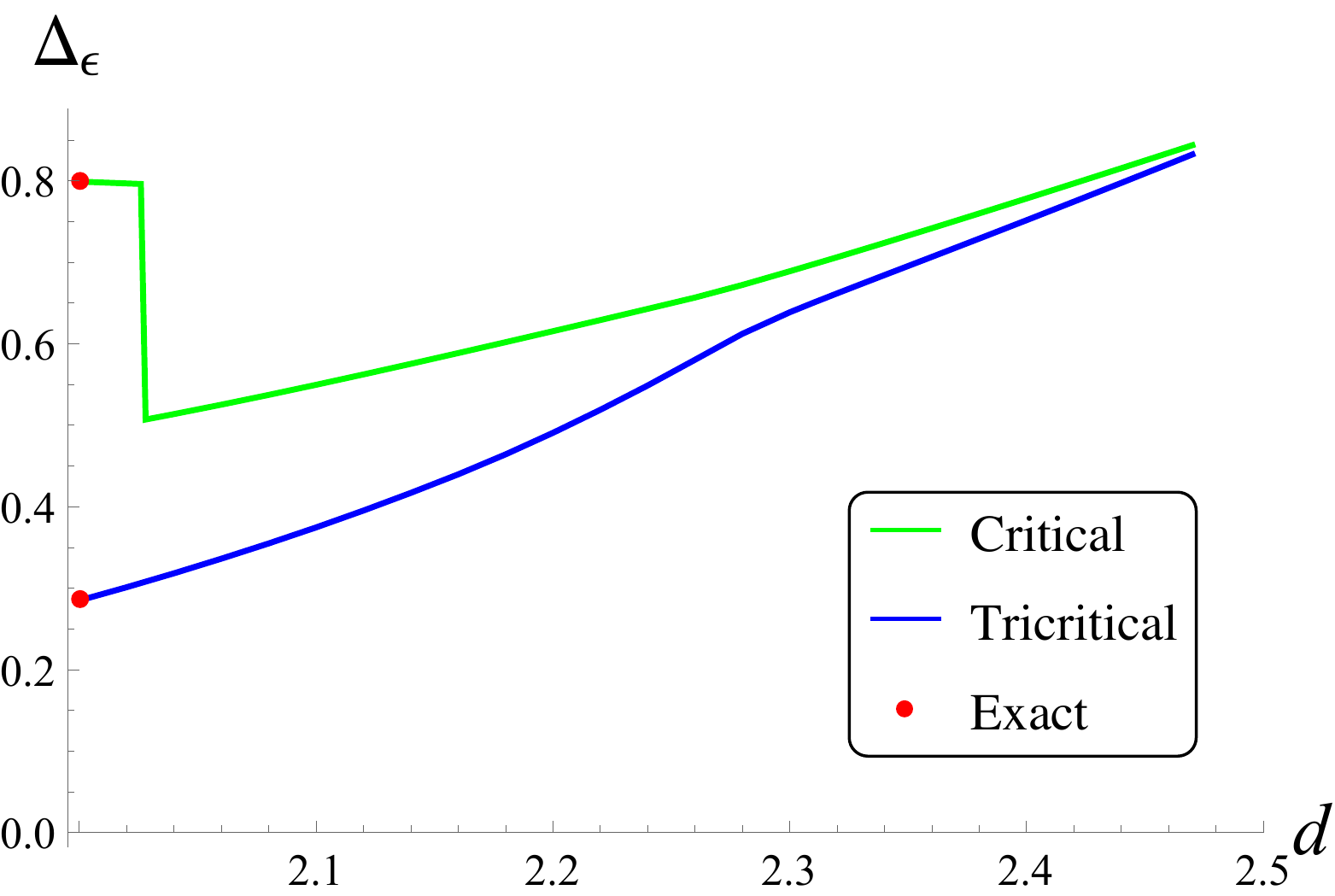}
	\includegraphics[width=\columnwidth]{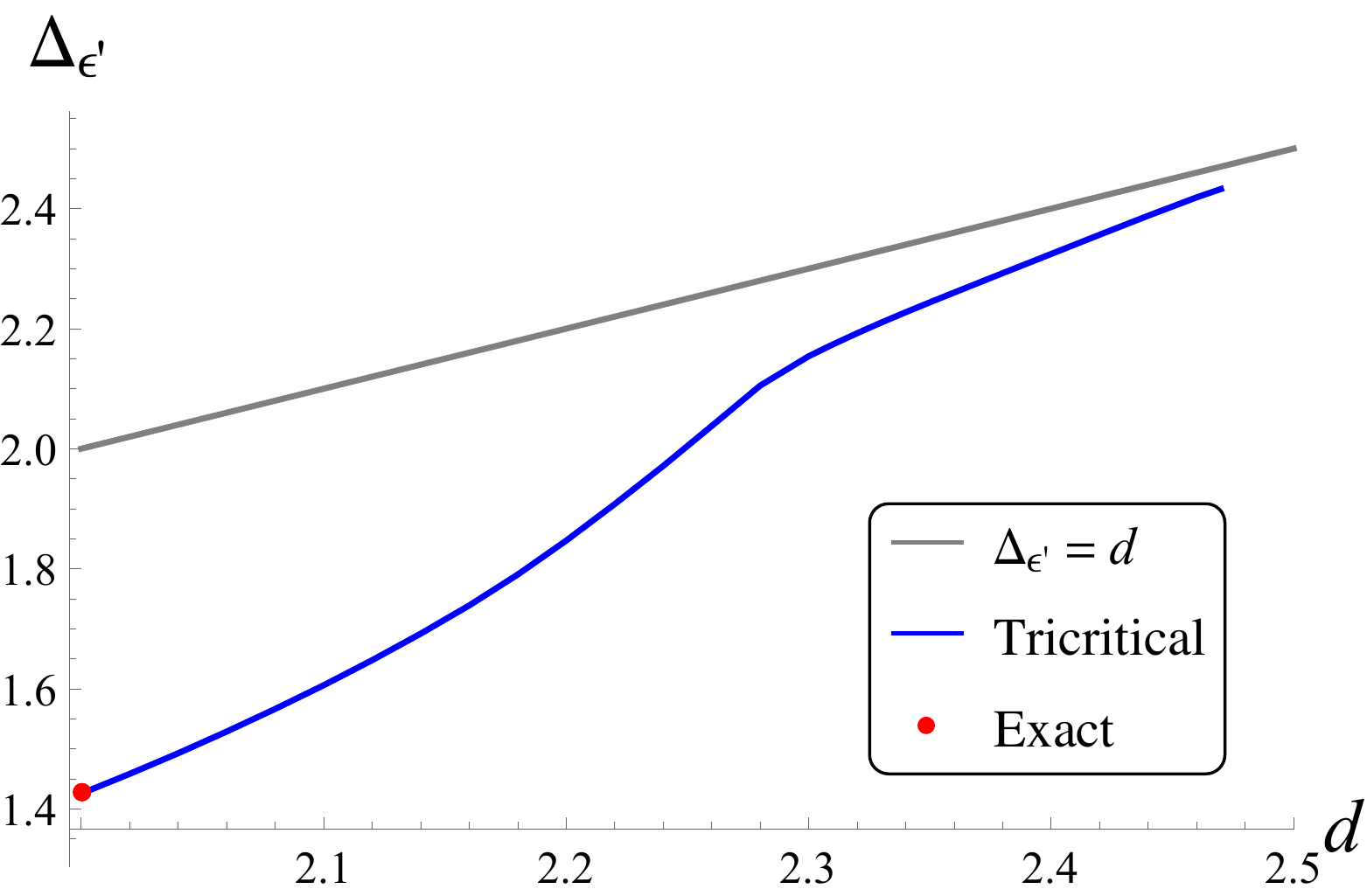}
	\caption{Scaling dimensions of relevant operators for the critical ({\bf green}) and tricritical ({\bf blue}) 3-state Potts models for $d\geq2$, computed using the bootstrap algorithm described in the main text. The red dots denote the exact 2d values \eqref{critVals} and \eqref{tricritVals}. The tricritical theory has an extra relevant singlet $\epsilon'$, whose scaling dimension we compare to the gray marginality line $\Delta_{\epsilon'}=d$.}
	\label{fig:spectrum}
\end{figure*}

We start by bootstrapping the critical theory by truncating the 39 crossing equations listed in \ref{configs}, rephrasing them as a semidefinite program as in \cite{Poland:2011ey}, which crucially assumes unitarity, and then solving these constraints efficiently using \texttt{SDPB} \cite{Simmons-Duffin:2015qma}. We assume that $\sigma, \sigma'$ are the only relevant scalars in the $\bf1$, $\epsilon$ is the only relevant scalar in the $\bf0_+$, and there are no relevant scalars in the $\bf0_-$. We also impose that these relevant operators are unique by scanning over the seven ratios of their OPE coefficients \footnote{Without the OPE scan, the bootstrap constraints would in principle allow for many operators with the same scaling dimension.}, as in \cite{Chester:2019ifh}. In Figure \ref{fig:2d}, we show the allowed region for $d=2$ in the space of $(\Delta_\sigma,\Delta_{\sigma'},\Delta_\epsilon)$, where we used bootstrap accuracy $\Lambda=11$ \footnote{As defined originally in \cite{Chester:2014fya}, this counts the number of derivatives that we use in the expansion of the conformal blocks around the crossing symmetric point. The bootstrap bounds shrink monotonically as a function of $\Lambda$.}. We find that the allowed region takes the shape of a cone, whose tip precisely matches the minimal model values \eqref{critMin}.

Instead of laboriously scanning over the full 10 parameters space of scaling dimensions and OPE ratios, we would like to find the tip of the cone as efficiently as possible. We can do this using the Navigator method \cite{Reehorst:2021ykw} \footnote{See \cite{Henriksson:2022gpa,Sirois:2022vth,Reehorst:2021hmp} for other uses of this method.}, which defines a function on the CFT data whose gradient quickly navigates the bootstrap algorithm to the boundary of the allowed region. Since we observe that the tip of the cone is given by the minimal value of $\Delta_{\sigma}$ along this boundary, we simply minimize the navigator function in terms of $\Delta_{\sigma}$ to find the critical theory (See Appendix \ref{sec:details} for more details). We find the values of $(\Delta_\sigma\,,\Delta_{\sigma'}\,,\Delta_{\epsilon})$ to be \footnote{The full result for the navigator for this run, as well as the others given in this work, are given in the attached \texttt{Mathematica} file.}
\es{critVals}{
d=2:\qquad \text{Navigator}:& \quad (.1332\,,1.332\,,.7991)\\
\text{Exact}:&\quad (.1333\,,1.333\,,.8)\,,
}
which compares well to the exact values in \eqref{critMin}. 

We can apply a similar algorithm to compute the tricritical theory. In this case we now allow for an extra relevant singlet $\epsilon'$. Since the search space is larger, for feasibility we restrict to the 21 crossing equations in \ref{configs} that only involve $\sigma$ and $\sigma'$. These correlators include nine different ratios of OPEs of the four relevant operators, which along with the relevant operator scaling dimensional makes for a 13-dimensional space. Using bootstrap accuracy $\Lambda=11$, we again minimize the navigator in terms of $\Delta_\sigma$ for $d=2$ to find the values of $(\Delta_\sigma\,,\Delta_{\sigma'}\,,\Delta_{\epsilon}\,,\Delta_{\epsilon'})$ to be
\es{tricritVals}{
d=2:\quad \text{Navigator}: &\quad (.0944\,, .947\,,.2848\,,1.425)\\
\text{Exact}:&\quad (.0952\,,.952\,,.285\,,1.428)\,,
}
which again compares well to the exact values in \eqref{tricritMin}.

Our algorithm for finding the critical and tricritical fixed point easily generalizes to $d>2$ \footnote{In fractional $d$, there is the possibility that some high twist operators might violate unitarity, as was shown explicitly for the Ising model in \cite{Hogervorst:2015akt}, which would invalidate the numerical bootstrap algorithm. On the other hand, these unitarity violating operators are typically of such high twist that they do not affect the numeral optimization of the truncated crossing equations, which are barely affected by high twist operators as shown in \cite{Pappadopulo:2012jk}. Indeed, in \cite{El-Showk:2013nia,Cappelli:2018vir,Chester:2015lej,Chester:2014gqa} bootstrap studies were carried out in fractional $d$ for both the Ising model and other universality classes, and the results in all cases matched independent predictions for the CFT.}, since the only dependence on $d$ is the conformal blocks in \eqref{4point}, and the definition of a relevant operator. In Figure \ref{fig:spectrum}, we show the results for both the critical and tricritical algorithms for $d>2$ \footnote{The plot for the critical theory shows a sharp discontinuity around $d\sim 2.03$, which may suggest that the curve to the right of that discontinuity is no longer related to the critical Potts theory. On the other hand, the tricritical curve is smooth all the way to $d\sim2.5$, and $\Delta_{\epsilon'}$ gets close to marginality just as the other scaling dimensions for each theory get close, which suggests that the critical Potts curve should also be trusted all the way to $d\sim2.5$.}. We see that the scaling dimensions get closer to each other until they seem to merge at around $d\sim 2.3$ for $\Delta_{\sigma}$, and $d\sim2.5$ for $\Delta_{\sigma'}$ and $\Delta_\epsilon$. For $\Delta_{\epsilon'}$ we only have access to the tricritical value, which approaches marginality also near $d\sim2.5$. We cannot yet observe the precise merger of the CFT data, because we found that the numerics became very unstable as we approach $d\sim2.5$, so in practice we just show results until $d=2.47$. We suspect this instability is caused by the near vicinity of of the two cone tips in the space of CFT data, which makes it difficult for the navigator to find each.

\section{Discussion}\label{sec:discussion}

In this work we found evidence for the upper critical dimension of the 3-state critical and tricritical Potts models. We used the recent Navigator bootstrap method to find kinks in the space of allowed critical exponents of each theory that for $d=2$ match the exactly known minimal models, and for $d>2$ get closer until around $d\sim2.5$ where they are about to merge. Our data suggests two possibilities for $d_\text{crit}$. The first is that we should identify $d_\text{crit}\sim2.5$ with the value where all data seems about to merge for each theory, and where $\Delta_{\epsilon'}$ goes to marginality as expected from the merger and annihilation scenario. From this perspective, the fact that $\Delta_\sigma$ seems to already merge at $d\sim2.3$ is just a curiosity, which does not contradict anything since $\Delta_\sigma$ for each theory remain close until $d\sim2.5$. It would be curious if $d_\text{crit}\sim2.5$, since its close to the exact value $d^6_\text{crit}=2.5$ for the pentacritical Ising model, as discussed around \eqref{dcriteasy}. In $d=2$ the pentacritical Ising model is the diagonal minimal model with the same central charge $c$ as the non-diagonal minimal model that describes the tricritical Potts model. It is possible that these two theories continue to share properties in $d>2$ that would explain the coincidence of $d_\text{crit}$.

The main difficulty with $d_\text{crit}\sim2.5$ is that the merger and annihilation scenario predicts that the operator going to marginality (i.e. $\Delta_{\epsilon'}$) should approach $d$ as \cite{Gorbenko:2018ncu}
\es{predict}{
\Delta_{\epsilon'}-d_\text{crit}\propto \sqrt{d}
} 
while the behavior in Figure \ref{fig:spectrum} seems linear near $d\sim2.5$. On the other hand, we do observe approximate square root behavior for $d\lesssim2.3$, which is also where $\Delta_\sigma$ merged and where the other scaling dimensions show kinks. The estimation $d_\text{crit}\sim2.3$ is also close to the value $(d_\text{crit},q_\text{crit})=(2.32,2.85)$ computed from lattice methods in \cite{PhysRevB.23.6055}, where $d_\text{crit}$ should decrease as $q_\text{crit}$ increases. If this interpretation is correct, then as we improve the bootstrap accuracy $\Lambda$ we would expect the kink in the $\Delta_{\epsilon'}$ plot near $d_\text{crit}\sim2.3$ to move upward toward the marginality line.

To determine which $d_\text{crit}$ is correct, we will likely need to improve $\Lambda$ far beyond the modest value of $\Lambda=11$ used in this work. This will require a drastic improvement of the current Navigator method, which we already pushed to its limits in this work. A higher $\Lambda$ bootstrap may even turn the kinks we observed into islands, so that our determination of the critical exponents will become completely rigorous. Nevertheless, since our estimate for the scaling dimensions moves up as we increase $\Lambda$\footnote{We checked this for various $d$ and $\Lambda$, and it can also be seen by comparing the exact values to our estimates in $d=2$ in \eqref{critVals} and \eqref{tricritVals}.}, and since we already observed $\Delta_{\epsilon'}$ going to marginality near $d\sim2.5$, we can already confidently bound the upper critical dimension as $d_\text{crit}\lesssim2.5$.

Looking ahead, it would also be nice to find the general critical curve $(d_\text{crit},q_\text{crit})$ for the $q$-state Potts model using the conformal bootstrap. One challenge is that for fractional $q$ we expect the theory to be logarithmic \cite{Gorbenko:2020xya,https://doi.org/10.48550/arxiv.2208.14298}, which strongly breaks unitarity so that the numerical bootstrap cannot be applied. This is unlike the possible breaking of unitarity for integer $q$ and fractional $d$, which is expected to be negligible \cite{Hogervorst:2015akt}.

Finally, the method we introduced in this work, of minimizing the navigator functional in terms of a certain critical exponent so as to find a kink associated with a known theory, could be useful to bootstrap other strongly coupled merger and annihilation scenarios. For instance, QED$_3$ with $N_f$ fermions is believed to stop being conformal below some $N_f^\text{crit}$. It has been difficult to precisely bootstrap this theory \cite{Chester:2017vdh,Chester:2016wrc,Li:2018lyb,Albayrak:2021xtd,He:2021xvg}. It is possible that a sharp signature for this theory could be found by bootstrapping a large set of correlators, which could be made feasible using our method.

\section*{Acknowledgments}

We thank Amnon Aharony, Ofer Aharony, David Huse, Yinchen He, Igor Klebanov, Walter Landry, Silvu Pufu, Junchen Rong, Leonardo Rastelli, Alessandro Vichi, and Bernardo Zan for useful conversations, and Silviu Pufu, Zhehan Qin, and Igor Klebanov for collaboration at an early stage of this project. SMC is supported by the Weizmann Senior Postdoctoral Fellowship, the Center for Mathematical Sciences and Applications and the Center for the Fundamental Laws of Nature at Harvard University. This project has received funding from the European Research Council (ERC) under the European Union’s Horizon 2020 research and innovation programme (grant agreement no. 758903). The authors would like to acknowledge the use of the WEXAC cluster in carrying out this work. We thank Yinchen He for support on computational resources. The computations in this paper were partially run on the Symmetry cluster of Perimeter institute.

\appendix

\section{Numerical bootstrap details}
\label{sec:details}

We used the \texttt{simpleboot} package \cite{simpleboot} for the bootstrap calculation in this paper. The semi-definite program is solved using \texttt{SDPB} \cite{Simmons-Duffin:2015qma,Landry:2019qug}. For all the computations, we used the following choice for the set of spins at each value of $\Lambda$:
\begin{align}\label{tab:spinsets}
S_{11} &= \{0,\dots,16\}. \nonumber
\end{align}
The pole order parameter is $\kappa=9$. For the navigator computations, we used Algorithm 2 in \cite{Reehorst:2021ykw} with $g_{\text{tol}}=10^{-15}$. The some of the \texttt{sdpb} parameters are given in table~\ref{tab:params}. We used the ``Feasiblity" column for the computation of Figure \ref{fig:2d}, and the ``Navigator" column for all the navigator computations.
\begin{table}
\begin{center}
\begin{tabular}{@{}|c|c|c|c|c@{}}
\hline
 &  \text{Feasibility} & \text{Navigator} \\
{\small\texttt{precision}} & 400 & 765 \\
{\small\texttt{dualityGapThreshold}} &  $10^{-20}$ & $10^{-30}$ \\
{\small\texttt{primalErrorThreshold}}&  $10^{-60}$ & $10^{-30}$ \\
{\small\texttt{dualErrorThreshold}} & $10^{-60}$ & $10^{-30}$ \\ 
{\small\texttt{maxComplementarity}} & $10^{100}$ & $10^{800}$\\
{\small\texttt{detectPrimalFeasibleJump}} & \text{true} & \text{false}\\
{\small\texttt{detectDualFeasibleJump}} & \text{true} & \text{false}\\
\hline
\end{tabular}
\caption{\label{tab:params}Parameters used for the navigator computations. }
\end{center}
\end{table}

To make the cone plot, using the cutting surface algorithm \cite{Chester:2019ifh}, we computed 383 points in the space of the scaling dimension. For each point, the average number of OPE coefficients sampled in the cutting surface algorithm was 90. The total computational time spent in \texttt{SDPB} was around 6800 CPU hours.
For the navigator runs in the Potts setup, the typical calculation for a given dimension took around 100 steps in the BFGS algorithm and the total time spent in \texttt{SDPB} was around 1100 CPU hours. For the navigator runs in the tricritical Potts setup, the typical calculation for a given dimension took around 170 steps in the BFGS algorithm and the total time spent in \texttt{SDPB} was around 1400 CPU hours. Those numbers depend on the choices of the initial point and initial Hessian for the BFGS algorithm. Once a calculation for some dimensions are finished, we made an extrapolation or interpolation for the initial data of the new dimension in order to accelerate the computation.

\onecolumngrid
\vspace{1in}
\twocolumngrid

\bibliographystyle{ssg}
\bibliography{potts_draft}

\end{document}